# The Fourth SM Family Neutrino at Future Linear Colliders

A. K. Ciftci[1], R. Ciftci[2], S. Sultansoy[2,3]

[1]Physics Department, Faculty of Sciences, Ankara University, 06100 Tandogan, Ankara, Turkey,

[2]Physics Department, Faculty of Sciences and Arts, Gazi University, 06500 Teknikokullar, Ankara, Turkey,

[3]Institute of Physics, Academy of Sciences, H. Cavid Avenue 33, Baku, Azerbaijan

## Abstract

It is known that Flavor Democracy favors the existence of the fourth standard model (SM) family. In order to give nonzero masses for the first three family fermions Flavor Democracy has to be slightly broken. A parametrization for democracy breaking, which gives the correct values for fundamental fermion masses and, at the same time, predicts quark and lepton CKM matrices in a good agreement with the experimental data, is proposed. The pair productions of the fourth SM family Dirac ($\nu_4$) and Majorana ($N_1$) neutrinos at future linear colliders with $\sqrt{s} = 500$ GeV, 1 TeV and 3 TeV are considered. The cross section for the process $e^+e^- \rightarrow \nu_4 \bar{\nu}_4 (N_1 N_1)$ and the branching ratios for possible decay modes of the both neutrinos are determined. The decays of the fourth family neutrinos into muon channels ($\nu_4(N_1) \rightarrow \mu^\pm W^\mp$) provide cleanest signature at $e^+e^-$ colliders. Meanwhile, in our parametrization this channel is dominant. W bosons produced in decays of the fourth family neutrinos will be seen in detector as either di-jets or isolated leptons. As an example we consider the production of 200 GeV mass fourth family neutrinos at $\sqrt{s} = 500$ GeV linear colliders by taking into account di-muon plus four-jet events as signatures.

# I. INTRODUCTION

It is well-known that standard model (SM) does not fix number of fermion families. There is only one indication that this number is less than 16 coming from asymptotic freedom of QCD. On the other hand, Flavor Democracy (or, in other words, the Democratic Mass Matrix) approach [1-4] favors the existence of the fourth SM family with the nearly degenerate masses in the range of 300-700 GeV [5-8]. Concerning the experimental situation: the LEPI data show that, there are three SM families with light neutrinos [9]. However extra SM families are allowed by the data, as long as the mass of new neutrinos are larger than $M_Z/2$. Furthermore presicion electroweak data does not exclude the fourth SM family, moreover two and even three extra generations are also allowed if $m_{v_4} \sim 50$ GeV [10-11]. Experimental constraints [9] on the masses of fundamental SM fermions are presented in Table I.

The fourth SM family quarks will be copiously produced at the LHC [12-13] if their masses are less than 1 TeV. The FNAL Tevatron Run II can observe $u_4$ and $d_4$ before the LHC if there is anomalous interaction with enough strength between the fourth family quarks and known quarks [14]. In addition, evidence for the extra SM families may come from the search for the SM Higgs boson due to an essential enhancement in the production of the Higgs boson via gluon-gluon fusion [15].

The observation of the fourth SM family leptons at hadron colliders is difficult due to a large background. Therefore, the fourth family leptons will be observed at lepton colliders with sufficient center of mass energy. This subject was investigated in [16] for muon colliders and in [17] for $e^+e^-$ and $\gamma\gamma$ colliders. In these papers the Dirac nature of the fourth SM family neutrino was assumed. Actually, the SM does not prohibit Majorana mass terms for right-handed neutrino. The fourth SM family Majorana neutrino search strategy changes greatly comparing to Dirac case.

In the four family case see-saw mechanism in principle, is not required to get light masses for the first three SM family neutrinos [18]. Meanwhile in the case of Majorana neutrinos, there will be double supression because of the both DMM and see-saw mechanism. The existence of the fourth family neutrinos leads to a number of cosmological consequences [19].

The most important barrier in front of high-energy electron-positron colliders is synchrotron radiation emitted by charged particles of circular motion. To avoid the resulting energy loss, one needs to build either a ring with a circumference of thousands of kilometres or a linear machine with the length of tens of kilometres. Because of the cost, only choice for the high energy colliders is the linear colliders.

The International Linear Collider (ILC), with the center of mass energy of 500 GeV (preferebly extendable to 1 TeV) and with $10^{34}$ cm$^{-2}$s$^{-1}$ luminosity, is being developed for use by particle physicists. The two technologies for the ILC use different types of cavities to accelerate electrons and positrons. The TESLA technology [20] has involved superconducting cavities operating at 2 K, whereas the technology of the NLC and JLC was based on copper cavities that would be run at room temperature. However, due to the huge cost of the linear collider the physicists selected only one. Following evaluation of limitations of each cavity type, the International Steering Committee preferred the superconducting approach. Assuming that the design work is completed on time, construction of the $\sqrt{s} = 0.5$ TeV machine could start about 2010. Meanwhile LHC will provide a first glimpse of any new physics at energies up to about 1 TeV. Therefore, depending on LHC results, a machine with higher energy than 1 TeV may be preferred. In this case, CLIC [21] will be right machine with the center of mass energy of 3 TeV and with $10^{35}$ cm$^{-2}$s$^{-1}$ luminosity. CLIC generates an accelerating gradient of 150 MV m$^{-1}$ with the resulting 20 km of active length. To reach this high accelerating gradient, CLIC uses two beam accelerator technology operating at 30 GHz radio frequency.

In this paper we consider pair production of the fourth SM family neutrinos at future $e^+e^-$ colliders. In section II basic assumptions of the flavor democracy hypothesis are given and the fourth family quark CKM matrix is evaluated. The leptonic sector is analysed in section III, where leptonic CKM matrix is reproduced by using the same parametrization for democracy breaking as in quark sector and possible decay modes of the fourth family Dirac and Majorana neutrinos are discussed. The numerical calculations for the processes $e^+e^- \to \nu_4 \bar{\nu}_4$ (Dirac case) and $e^+e^- \to N_1 N_1$ (Majorana case) are performed in section IV using the Comphep 4.4.3 package. Finally, we give some concluding remarks in section V.

## II. FLAVOR DEMOCRACY AND THE FOURTH SM FAMILY

It is useful to consider three different bases:
- Standard model basis $\{f^0\}$,
- Mass basis $\{f^m\}$ and

- Weak basis $\{f^w\}$.

According to the three-family SM, before the spontaneous symmetry breaking quarks are grouped into the following SU(2)×U(1) multiplets:

$$\begin{pmatrix} u_L^0 \\ d_L^0 \end{pmatrix}, u_R^0, d_R^0; \begin{pmatrix} c_L^0 \\ s_L^0 \end{pmatrix}, c_R^0, s_R^0; \begin{pmatrix} t_L^0 \\ b_L^0 \end{pmatrix}, b_R^0, t_R^0. \tag{1}$$

In one-family case all bases are equal and for example, d-quark mass is obtained due to Yukawa interaction

$$L_Y^{(d)} = a_d \begin{pmatrix} \overline{u}_L & \overline{d}_L \end{pmatrix} \begin{pmatrix} \varphi^+ \\ \varphi^- \end{pmatrix} d_R + h.c. \Rightarrow L_m^{(d)} = m_d \overline{d}d \tag{2}$$

where $m_d = a_d \eta$, $\eta = \langle \varphi^0 \rangle \cong 249$ GeV. In the same manner $m_u = a_u \eta$, $m_e = a_e \eta$ and $m_{\nu_e} = a_{\nu_e} \eta$ (if neutrino is Dirac particle). In n-family case

$$L_Y^{(d)} = \sum_{i,j=1}^n a_{ij}^d \begin{pmatrix} \overline{u}_{Li}^0 & \overline{d}_{Li}^0 \end{pmatrix} \begin{pmatrix} \varphi^+ \\ \varphi^- \end{pmatrix} d_{Rj}^0 + h.c. \Rightarrow L_m^{(d)} = \sum_{i,j=1}^n m_{i,j}^d \overline{d}_i^0 d_j^0, \quad m_{i,j}^d = a_{ij}^d \eta \tag{3}$$

where $d_1^0$ denotes $d^0$, $d_2^0$ denotes $s^0$ etc. The diagonalization of mass matrix of each type of fermions, or in other words transition from SM basis to mass basis, is performed by well-known bi-unitary transformation. Then, the transition from mass basis to weak basis result in CKM matrix

$$U_{CKM} = (U_L^u)^+ U_L^d \tag{4}$$

which contains 3 (6) observable mixing angles and 1 (3) observable CP-violating phases in the case of three (four) SM families. Before the spontaneous symmetry breaking, all quarks are massless and three are no differences between $d^0$, $s^0$ and $b^0$. In other words, fermions with the same quantum numbers are indistinguishable. This leads us to the **first assumption** [1-3, 22], namely, Yukawa couplings are equal within each type of fermions:

$$a_{ij}^d \cong a^d, \ a_{ij}^u \cong a^u, \ a_{ij}^l \cong a^l, \ a_{ij}^\nu \cong a^\nu \tag{5}$$

The first assumption result in n-1 massless particles and one massive particle with $m = na^F \eta$ (F = u, d, l, ν) for each type of the SM fermions.

Because there is only one Higgs doublet which gives Dirac masses to all four types of fermions, it seems natural to make the **second assumption** [5,6], namely, Yukawa constant for different types of fermions should naturally be equal:

$$a^d \approx a^u \approx a^l \approx a^\nu. \qquad (6)$$

Taking into account the mass values for the third generation, the second assumption leads to the statement that according to the flavor democracy, the fourth SM family should exist.

In terms of the mass matrix, the above arguments mean

$$M^0 = a\eta \begin{pmatrix} 1 & 1 & 1 & 1 \\ 1 & 1 & 1 & 1 \\ 1 & 1 & 1 & 1 \\ 1 & 1 & 1 & 1 \end{pmatrix} \Rightarrow M^m = 4a\eta \begin{pmatrix} 0 & 0 & 0 & 0 \\ 0 & 0 & 0 & 0 \\ 0 & 0 & 0 & 0 \\ 0 & 0 & 0 & 1 \end{pmatrix} \qquad (7)$$

Now, let us make the **third assumption**, namely, a is between $e = g_W \sin\theta_W = \sqrt{4\pi\alpha_{em}}$ and $g_Z = g_W / \cos\theta_W$. Therefore, the fourth family fermions are almost degenerate, in good agreement with experimental value $\rho$ = 0.9998±0.0008 [23], and their common mass lies between 320 GeV and 730 GeV. The last value is close to upper limit on heavy quark masses, $m_Q \leq 700\,\text{GeV}$, which follows from partial-wave unitary at high energies [24]. It is interest that with preferable value $a \approx g_W$, flavor democracy predicts $m_4 \approx 8 m_W \approx 640$ GeV.

In order to give nonzero masses for the first three SM family fermions flavor democracy has to be slightly broken [7]. To perform this scheme one has to consider to get the masses and CKM mixing matrix elements in the correct experimental range. Below we use following parametrization for democracy breaking (assuming a modification which has a minimum effect on full democracy):

$$M^0 = a\eta \begin{pmatrix} 1 & 1+\gamma & 1+\beta & 1-\beta \\ 1+\gamma & 1+2\gamma & 1+\beta & 1-\beta \\ 1+\beta & 1+\beta & 1+\alpha & 1-\alpha \\ 1-\beta & 1-\beta & 1-\alpha & 1+\alpha+2\beta \end{pmatrix} \qquad (8)$$

At the limit of $\gamma = \beta = 0$ this matrix becomes the one given in [6].

Current limits [9] on the known quark masses are presented in the Table I, where the renormalization scale has been chosen to be $\mu$ = 2 GeV for light quarks (q = u, d, s) and $\mu = m_q$ for heavy quarks (q = c, b, t). At the electroweak scale ($\mu \cong m_Z$), the mass values are converted into the ones given in the Table II following the procedure presented in [25]. Eigenvalues of matrix (3) give us masses of corresponding fermions which are used to fix the values of parameters α, β and γ. In Tables III and IV we present these values for the up- and down-quark sectors with predicted values of the fourth family quark masses, taking g equal to $g_W$ and e, respectively. The fourth SM family quarks' mass values $m_{q_4}(\mu \cong m_Z) \approx 400$ GeV correspond to $m_{q_4}(\mu = m_{q_4}) \approx 320$ GeV, and $m_{q_4}(\mu \cong m_Z) \approx 800$ GeV to $m_{q_4}(\mu = m_{q_4}) \approx 640$ GeV.

The quark CKM matrix is given as $O_{CKM} = O_u O_d^T$, where $O_u$ and $O_d$ are (real) rotations which diagonalize up- and down-quarks mass matrices. (We assume that 3 phase parameters in the quarks' CKM matrix are small enough to be neglected.) With the parameters given in Table III, one obtains

$$O_{CKM} = \begin{pmatrix} 0.9747 & -0.2235 & -0.0028 & -0.0001 \\ 0.2232 & -0.9738 & -0.0439 & -0.0006 \\ -0.0125 & 0.0422 & -0.9990 & -0.0008 \\ -0.0002 & 0.0005 & 0.0008 & -1.0000 \end{pmatrix} \quad (9)$$

With the parameters given in Table IV, the CKM matrix of quarks takes form

$$U_{CKM} = \begin{pmatrix} 0.9747 & 0.2236 & -0.0030 & -0.0002 \\ 0.2232 & -0.9738 & -0.0439 & -0.0012 \\ -0.0125 & -0.0422 & 0.9990 & -0.0014 \\ 0.0005 & -0.0011 & -0.0014 & 1.0000 \end{pmatrix} \quad (10)$$

These matrices should be compared with the experimental one

$$\begin{pmatrix} 0.9730 - 0.9746 & 0.2174 - 0.2241 & 0.0030 - 0.0044 & * \\ 0.213 - 0.226 & 0.968 - 0.975 & 0.039 - 0.044 & * \\ 0 - 0.08 & 0 - 0.11 & 0.07 - 0.9993 & * \\ * & * & * & * \end{pmatrix} \quad (11)$$

taken from the Review of Particle Physics [9]. It is seen that our predictions are in good agreement with experimental data.

**III. THE FOURTH SM FAMILY NEUTRINO:**

**A. Dirac case**

In the leptonic sector, we know masses of charged leptons precisely, whereas experiments give only upper limits for neutrino masses. In Tables V and VI we present α, β, γ parameters and corresponding masses for the leptonic sector with predicted mass values of the fourth SM family leptons taking a equal to $g_W$ and e, respectively. In our predictions known neutrino masses are almost degenerate and the squared-mass difference are $\Delta m_{SUN}^2 = \Delta m_{21}^2 = 1.7 \times 10^{-4} (eV)^2$ and $\Delta m_{ATM}^2 = \Delta m_{31}^2 = 4.95 \times 10^{-3} (eV)^2$ which should be compared with the experimental data $1.2 \times 10^{-4} (eV)^2 < \Delta m_{SUN}^2 < 1.9 \times 10^{-4} (eV)^2$ and $1.4 \times 10^{-3} (eV)^2 < \Delta m_{ATM}^2 < 5.1 \times 10^{-3} (eV)^2$ [26].

The leptonic CKM matrix is $O^l_{CKM} = O_\nu O_l^T$, where $O_\nu$ and $O_l$ are rotations which diagonalize neutrino and charged lepton mass matrices. With the parameters given in Table V, one obtains

$$U^l_{CKM} = \begin{pmatrix} 0.86 & -0.42 & -0.29 & 7.92 \cdot 10^{-5} \\ -0.51 & -0.65 & -0.56 & 10.85 \cdot 10^{-5} \\ 0.05 & 0.63 & -0.77 & 7.34 \cdot 10^{-5} \\ -1.55 \cdot 10^{-5} & 5.74 \cdot 10^{-5} & 1.41 \cdot 10^{-4} & 1.00 \end{pmatrix} \quad (12)$$

With the parameters given in Table VI, the CKM matrix of leptons takes form

$$U^l_{CKM} = \begin{pmatrix} 0.86 & -0.42 & -0.29 & 1.58 \cdot 10^{-4} \\ -0.51 & -0.65 & -0.56 & 2.17 \cdot 10^{-4} \\ 0.05 & 0.63 & -0.77 & 1.46 \cdot 10^{-4} \\ -3.11 \cdot 10^{-5} & 1.15 \cdot 10^{-4} & 2.81 \cdot 10^{-4} & 1.00 \end{pmatrix} \quad (13)$$

These matrices should be compared with the experimental matrix

$$\begin{pmatrix} 0.71-0.88 & 0.46-0.68 & 0.00-0.22 \\ 0.08-0.66 & 0.26-0.79 & 0.55-0.85 \\ 0.10-0.66 & 0.28-0.80 & 0.51-0.83 \end{pmatrix} \quad (14)$$

taken from [26]. As can be seen, our predictions are roughly in agreement with experimental data. Note that the values in Eq.(14), which are estimated for three neutrino case, might be relaxed in four neutrino case (as it happens in quark sector [9]).

With predicted fourth family lepton masses, given in Tables V and VI and lepton CKM matrices (7) and (8), one sees that the decay modes of the fourth SM family neutrinos are following: $Br(\nu_4 \to e^- + W^+) \approx 0.27$, $Br(\nu_4 \to \mu^- + W^+) \approx 0.5$, $Br(\nu_4 \to \tau^- + W^+) \approx 0.23$.

**B. Majorana case**

As mentioned above, the SM does not prohibit Majorana mass terms for right-handed neutrinos. Therefore, $(4 \times 4)$ mass matrix is replaced by $(8 \times 8)$ mass matrix:

$$\frac{1}{2} \begin{pmatrix} \overline{\nu_L^0}^i & \overline{\nu_R^0}^i \end{pmatrix} \begin{pmatrix} 0 & m^\nu_{ij} \\ m^\nu_{ij} & M_{ij} \end{pmatrix} \begin{pmatrix} \nu_L^{0\,j} \\ \nu_R^{0\,j} \end{pmatrix} \quad (15)$$

where i, j = 1, 2, 3, 4 (in this subsection we follow notations of Ref. [27]).

According to the flavor democracy $m^\nu_{ij} = a^\nu_{ij}\eta = a^\nu \eta = a\eta$ and $M_{ij} = M$, where M is the Majorana mass scale of right-handed neutrinos. As a result of transition from SM basis to mass basis we obtain six massless Majorana neutrinos and two massive Majorana neutrinos with

$m_1 = 2\left(\sqrt{4(a\eta)^2 + M^2} - M\right)$ and $m_2 = 2\left(\sqrt{4(a\eta)^2 + M^2} + M\right)$. In this case, while breaking flavor democracy, one should keep contributions to known neutrinos ($\nu_e$, $\nu_\mu$, $\nu_\tau$) from other neutrino components small enough in order to avoid contradictions with experimental data on the weak charged currents.

The last condition can be satisfied naturally if the Majorana mass terms has the form $M_{ij} = M\delta_{ij}$. This assumes that an additional discrete symmetry takes place for right-handed neutrinos. In this case transition from SM basis to mass basis leads to mass eigenvalues of $m = \{0,0,0,m_1,M,M,M,m_2\}$ where $m_1 = 2\left(\sqrt{64(a\eta)^2 + M^2} - M\right)$ and $m_2 = 2\left(\sqrt{64(a\eta)^2 + M^2} + M\right)$. The situation considered in Ref. [27] corresponds to $m_{ij}^\nu = m_i^\nu \delta_{ij}$ and $M_{ij} = M\delta_{ij}$ which leads to usual see-saw mechanism.

In this paper we deal with properties of the fourth SM family neutrinos. We have assumed that the light fourth family Majorana neutrino mixes with known neutrinos in the same manner as Dirac case. Let us consider the fourth family neutrinos in details. Defining $N_1$ and $N_2$ as light and heavy mass eigenstates of the fourth family Majorana neutrinos, we have

$$N_1 = \cos\theta\, \nu_{4L}^0 - \sin\theta\, \nu_{4R}^0 \tag{16a}$$

$$N_2 = \sin\theta\, \nu_{4L}^0 + \cos\theta\, \nu_{4R}^0 \tag{16b}$$

with corresponding mass eigenvalues

$$m_{1,2} = 2\left(\sqrt{M^2 + 4(a\eta)^2} \mp M\right) \tag{17}$$

where $\tan 2\theta = (a\eta)/M$.

In Table VII and VIII we present the estimated values of heavy fourth family Majorana neutrino mass and mixing angle for various $m_1$ values with $a = e$ and $a = g_W$, respectively. Right-handed Majorana mass scale M is assumed to be larger than Dirac case $a\eta$, which corresponds to $\tan 2\theta < 1$. Therefore, upper limits on $m_1$ are 200 GeV and 400 GeV for $a = e$ and $a = g_W$, respectively.

It is seen that the ILC with with $\sqrt{s} = 500$ GeV permits only pair production of $N_1$, whereas $N_1N_2$ and $N_2N_2$ production could be possible at higher center of mass energies. However, corresponding cross-sections are supressed by factors of $\sin^2\theta$ for $N_1N_2$ and $\sin^4\theta$ for $N_2N_2$ in addition to kinematical suppression. Moreover, the dominant decay mode of $N_2$ will be $N_2 \to l_4^{\pm} W^{\mp}$ since $m_{l_4} < m_2$; with subsequent decay $l_4 \to N_1 W$. For these reasons we will focus our attention on the process $e^+e^- \to N_1 N_1$.

The part of the interaction Lagrangian responsible for production and decays of $N_1$ follows:

$$L = -\frac{g_W}{4\cos\theta_W}\cos^2\theta \overline{N_1}\gamma^\mu \gamma^5 N_1 Z_\mu - \left[\sum_{i=1}^{3}\frac{g_W}{2\sqrt{2}}\cos\theta O_{4i}\overline{l_i}\gamma^\mu(1-\gamma^5)N_1 W_\mu^- + h.c.\right] \quad (18)$$

where $l_1 = e$, $l_2 = \mu$ and $l_3 = \tau$. For numerical calculations we have used the $O_{4i}$ values given in Equations (12) and (13). As a result, estimated branching ratios are following: Br($N_1 \to e^- + W^+$) = Br($N_1 \to e^+ + W^-$) ≈ 0.135, Br($N_1 \to \mu^- + W^+$) = Br($N_1 \to \mu^+ + W^-$) ≈ 0.25 and Br($N_1 \to \tau^- + W^+$) = Br($N_1 \to \tau^+ + W^-$) ≈ 0.115.

## IV. PAIR PRODUCTION OF THE FOURTH SM FAMILY NEUTRINO AT E⁺E⁻ COLLIDERS

In this section we analyse the processes $e^+e^- \to N_1 N_1$ and $e^+e^- \to \nu_4 \overline{\nu}_4$. For numerical calculations we have implemented the fourth SM family leptons into the CompHEP 4.4.3 package [28]. The computed cross sections as a function of neutrino masses at three different center of mass energies, namely $\sqrt{s} = 0.5$ TeV (ILC), $\sqrt{s} = 1$ TeV (ILC or CLIC) and $\sqrt{s} = 3$ TeV (CLIC), are given in Figures 1, 2 and 3, respectively. Following arguments given in the previous section, we cut short the mass of $m_1$ at 200 GeV for a = e and 400 GeV for a = $g_W$. Low value of Majorana neutrino production cross section with respect Dirac neutrino one originates from two points: kinematical supression [29,30] and mixing angle θ (see Tables VII and VIII). Indeed the ratio of cross sections for Majorana and Dirac neutrinos is given by

$$\frac{\sigma(e^+e^- \to N_1 N_1)}{\sigma(e^+e^- \to \nu_4 \overline{\nu}_4)} = \frac{4\beta \cos^4\theta}{3+\beta^2} \quad (19)$$

where $\beta = \left[1-(2m/s)^2\right]^{1/2}$ and cosθ is defined in Equations (16a) and (16b). The expected event numbers per year for a several mass values are presented in Table IX.

The decays of the fourth family neutrinos into muon channels provide cleanest signature at $e^+e^-$ colliders. Meanwhile, in our parametrization this channel is dominant. In Majorana case same sign di-muons signature does not have any background. The total number of ($\mu^+\mu^+W^-W^-$) and ($\mu^-\mu^-W^+W^+$) events is 1/8 of the values given in Table IX. In Dirac case di-muon channel results in ($\mu^+\mu^-W^+W^-$) events and their number is quarter of the values given in Table IX. The background from SM with three families computed by using CompHEP leads to 830 events for $\sqrt{s} = 0.5$ TeV with 100 fb$^{-1}$, 1600 events for $\sqrt{s} = 1$ TeV with 300 fb$^{-1}$ and 2000 events for $\sqrt{s} = 3$ TeV with 1000 fb$^{-1}$.

W bosons produced in decays of the fourth family neutrinos will be seen in detector as either di-jets or isolated leptons. Taking in mind the reconstruction of the fourth family neutrino mass we assume that at least one W boson is decaying into di-jet. Excluding the final states containing τ leptons, the expected event topologies are presented in Table X. As it is seen from the table the events with clean signature constitute about half of the total number of events.

As an example we would like to consider production of 200 GeV mass fourth family neutrinos at $\sqrt{s} = 500$ GeV linear collider by taking into account di-muon plus four jet events as signatures. In the Dirac case 610 signal and 380 background events are expected. Concerning the fourth family Majorana neutrino the events with the same sign di-muons topology do not have significant SM background. In this case we expect 100 (130) signal events for a = e (a = $g_W$).

## V. CONCLUSION

Future lepton colliders will give a clear answer on the question whether the fourth family neutrino Dirac nature or Majorana one is. The clearest signature for Majorana neutrino case will be provided by same-sign dileptons accompaning with either four jets or $2j + l + \not{P}_T$. In Dirac case channel with the opposite-sign dileptons accompaning with four jets seems to be preferable one. It is known that the number of signal events with these topologies are sufficiently high to investigate the fourth family neutrino properties in details.

This work is supported in part by Turkish Planning Organization (DPT) under the Grant No 2002K120250.

**Table I.** Masses of fundamental SM fermions in units $GeV/c^2$[9].

| Neutrinos | Charged leptons | Up quarks | Down quarks |
|---|---|---|---|
| $\nu_e$: <3×10$^{-9}$ | e: 0.51099890×10$^{-3}$ | u: (1.5-4.0)×10$^{-3}$ | d: (4-8)×10$^{-3}$ |
| $\nu_\mu$: <0.19×10$^{-3}$ | μ: 0.105658357 | c: 1.15-1.35 | s: (80-130)×10$^{-3}$ |
| $\nu_\tau$: <18.2×10$^{-3}$ | τ: 1.77699 | t: 174.3±5.1 | b: 4.1-4.4 |
| $\nu_4$: >45 (stable) $\nu_4$: >90.3 (unstable) | $l_4$: >102.6 (stable) $l_4$: >100.8 (unstable) | $u_4$: >200 | $d_4$: >128 (charged current decay) $d_4$: >199 (neutral current decay) |

**Table II.** Masses of known quarks at $\mu \cong m_Z$ scale.

| Up quarks | $m_u$: (0.92-2.75) MeV | $m_c$: (545-763) MeV | $m_t$: 184.4±5.4 GeV |
|---|---|---|---|
| Down quarks | $m_d$: (3.06-5.20) MeV | $m_s$: (48.9-94.8) MeV | $m_b$: (2.82-3.17) GeV |

**Table III.** Parameters and corresponding mass values for quark sector (at $\mu \cong m_Z$) taking a = $g_W$.

| | γ = -0.00024 | β = -0.005424 | α = 0.464 | |
|---|---|---|---|---|
| Up quarks | $m_u$ = 2.03 MeV | $m_c$ = 564.3 MeV | $m_t$ = 186.714 GeV | $m_{u_4}$ = 799.411 GeV |
| | γ = 0.0001016 | β = 0.0002152 | α = 0.0072 | |
| Down quarks | $m_d$ = 4.21 MeV | $m_s$ = 48.94 MeV | $m_b$ = 2.84 GeV | $m_{d_4}$ = 800.042 GeV |

**Table IV.** Parameters and corresponding mass values for quark sector (at $\mu \cong m_Z$) taking a = e.

| | γ = -0.00048 | β = -0.10848 | α = 0.928 | |
|---|---|---|---|---|
| Up quarks | $m_u$ = 2.03 MeV | $m_c$ = 564.3 MeV | $m_t$ = 186.71 GeV | $m_{u_4}$ = 399.41 GeV |
| | γ = 0.0002032 | β = 0.0004304 | α = 0.0144 | |
| Down quarks | $m_d$ = 4.2 MeV | $m_s$ = 48.94 MeV | $m_b$ = 2.84 GeV | $m_{d_4}$ = 400.042 GeV |

**TableV.** Parameters and corresponding mass values for lepton sector taking a = $g_W$.

| Charged leptons | $\gamma = 6.557\times10^{-5}$ | $\beta = 7.474\times10^{-4}$ | $\alpha = -5.817\times10^{-3}$ | |
|---|---|---|---|---|
| | $m_e = 0.511$ MeV | $m_\mu = 105.634$ MeV | $m_\tau = 1.777$ GeV | $m_{l_4} = 640.07$ GeV |
| Neutrinos | $\gamma = -1.1\times10^{-13}$ | $\beta = -0.8\times10^{-13}$ | $\alpha = -1.3\times10^{-13}$ | |
| | $m_{\nu_e} = 0.780\times10^{-2}$ eV | $m_{\nu_\mu} = 1.528\times10^{-2}$ eV | $m_{\nu_\tau} = 7.083\times10^{-2}$ eV | $m_{\nu_4} = 640$ GeV |

**TableVI.** Parameters and corresponding mass values for lepton sector taking a = e.

| Charged leptons | $\gamma = 1.3114\times10^{-4}$ | $\beta = 1.4948\times10^{-3}$ | $\alpha = -1.1634\times10^{-2}$ | |
|---|---|---|---|---|
| | $m_e = 0.511$ MeV | $m_\mu = 105.632$ MeV | $m_\tau = 1.777$ GeV | $m_{l_4} = 320.07$ GeV |
| Neutrinos | $\gamma = -2.2\times10^{-13}$ | $\beta = -1.6\times10^{-13}$ | $\alpha = -2.6\times10^{-13}$ | |
| | $m_{\nu_e} = 0.780\times10^{-2}$ eV | $m_{\nu_\mu} = 1.528\times10^{-2}$ eV | $m_{\nu_\tau} = 7.083\times10^{-2}$ eV | $m_{\nu_4} = 320$ GeV |

**TableVII.** The estimated values of heavy fourth family Majorana neutrino mass and mixing angle for a = e.

| $M_1$, GeV | 50 | 75 | 100 | 125 | 150 | 175 | 200 |
|---|---|---|---|---|---|---|---|
| $M_2$, GeV | 2048 | 1365 | 1024 | 819 | 683 | 585 | 512 |
| $\cos\theta$ | 0.9968 | 0.9926 | 0.9861 | 0.9767 | 0.9637 | 0.9456 | 0.9214 |

**TableVIII.** The estimated values of heavy fourth family Majorana neutrino mass and mixing angle for a = $g_W$.

| $M_1$, GeV | 50 | 100 | 150 | 200 | 250 | 300 | 350 | 400 |
|---|---|---|---|---|---|---|---|---|
| $M_2$, GeV | 8192 | 4096 | 2731 | 2048 | 1683 | 1365 | 1170 | 1024 |
| $\cos\theta$ | 0.9992 | 0.9968 | 0.9926 | 0.9861 | 0.9767 | 0.9636 | 0.9456 | 0.9214 |

**Table IX.** Numbers of produced neutrino pairs in a working year for different center of mass energies.

| | $\sqrt{s} = 0.5$ TeV $L = 100$ fb$^{-1}$ | | | $\sqrt{s} = 1$ TeV $L = 300$ fb$^{-1}$ | | | $\sqrt{s} = 3$ TeV $L = 1000$ fb$^{-1}$ | | |
|---|---|---|---|---|---|---|---|---|---|
| | $N_1 N_1$ | | | $N_1 N_1$ | | | $N_1 N_1$ | | |
| m (GeV) | a = e | a = $g_W$ | $\nu_4 \bar{\nu}_4$ | a = e | a = $g_W$ | $\nu_4 \bar{\nu}_4$ | a = e | a = $g_W$ | $\nu_4 \bar{\nu}_4$ |
| 100 | 7700 | 8000 | 9300 | 6700 | 7000 | 7300 | 2600 | 2700 | 2750 |
| 150 | 4700 | 5300 | 7700 | 5600 | 6400 | 7000 | 2300 | 2600 | 2740 |
| 200 | 1700 | 2200 | 5300 | 4200 | 5500 | 6700 | 1900 | 2500 | 2720 |
| 250 | - | - | - | - | 4500 | 6100 | - | 2400 | 2700 |
| 300 | - | - | - | - | 3300 | 5500 | - | 2200 | 2680 |
| 350 | - | - | - | - | 2200 | 4700 | - | 2000 | 2650 |
| 400 | - | - | - | - | 1200 | 3800 | - | 1800 | 2600 |

**Table X.** Expected event topologies.

| Events | Branching ratios (%) | |
|---|---|---|
| | $N_1 N_1$ | $\nu_4 \bar{\nu}_4$ |
| $\mu^+ \mu^+ (\mu^- \mu^-) + 4j$ | 2.89 (2.89) | - |
| $\mu^+ \mu^- + 4j$ | 5.78 | 11.56 |
| $\mu^+ \mu^+ (\mu^- \mu^-) + 2j + l + \not{P}_T$ | 1.87 (1.87) | - |
| $\mu^+ \mu^- + 2j + l + \not{P}_T$ | 3.74 | 7.48 |
| $e^+ e^+ (e^- e^-) + 4j$ | 0.84 (0.84) | - |
| $e^+ e^- + 4j$ | 1.68 | 3.37 |
| $e^+ e^+ (e^- e^-) + 2j + l + \not{P}_T$ | 0.54 (0.54) | - |
| $e^+ e^- + 2j + l + \not{P}_T$ | 1.09 | 2.18 |
| $\mu^+ e^+ (\mu^- e^-) + 4j$ | 3.12 (3.12) | - |
| $\mu^+ e^- (\mu^- e^+) + 4j$ | 3.12 (3.12) | 6.24 (6.24) |
| $\mu^+ e^+ (\mu^- e^-) + 2j + l + \not{P}_T$ | 2.02 (2.02) | - |
| $\mu^+ e^- (\mu^- e^+) + 2j + l + \not{P}_T$ | 2.02 (2.02) | 4.04 (4.04) |

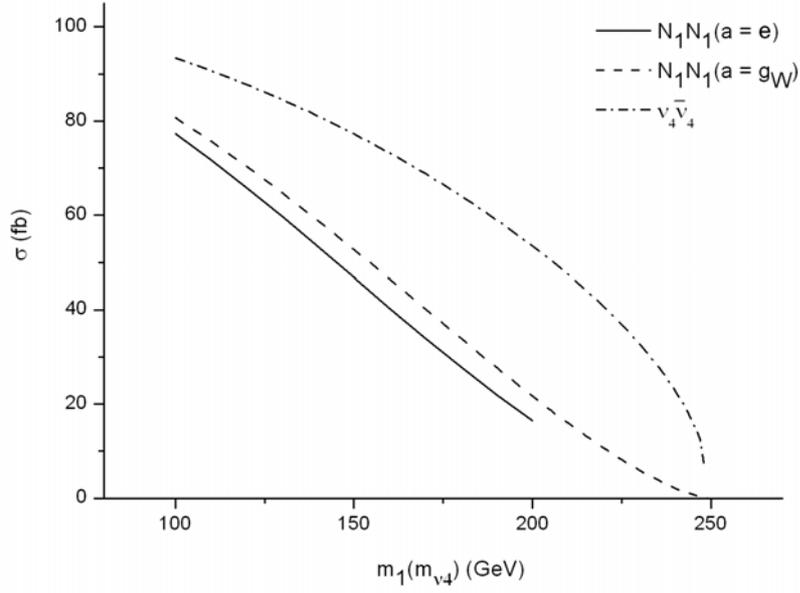

**Fig. 1.** Cross section for pair production of the fourth family neutrinos at $\sqrt{s} = 500$ GeV.

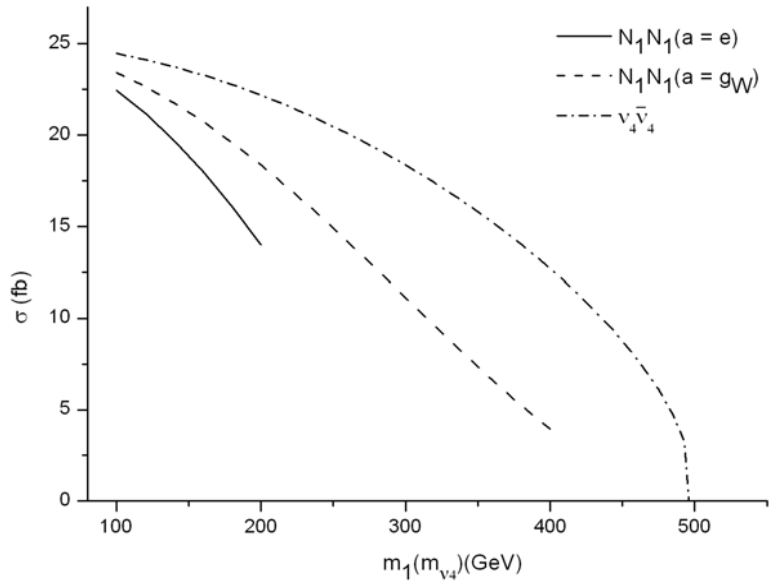

**Fig. 2.** Cross section for pair production of the fourth family neutrinos at $\sqrt{s} = 1$ TeV.

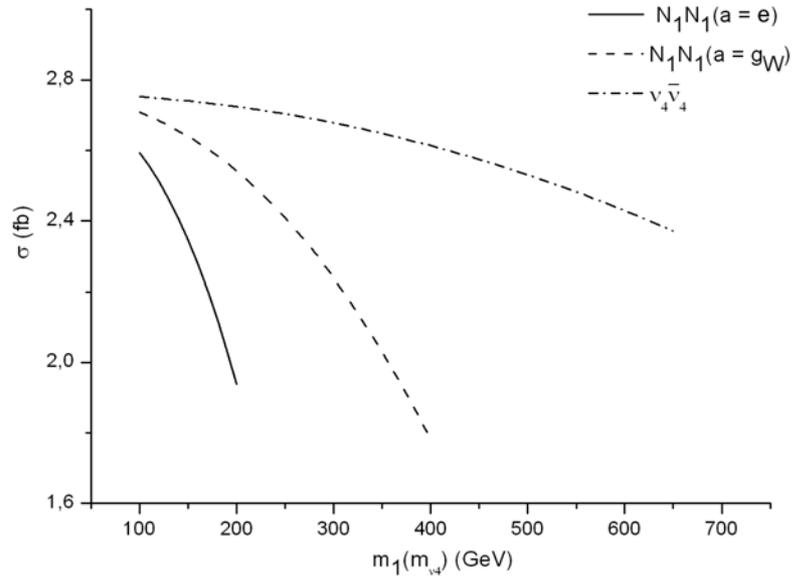

**Fig. 3.** Cross section for pair production of the fourth family neutrinos at $\sqrt{s} = 3$ TeV.